\documentclass[prb,reprint,superscriptaddress,twocolumn]{revtex4-1}

\usepackage{tabularx,booktabs}
\usepackage{multirow}
\usepackage{amsmath}
\usepackage{amssymb}
\usepackage{graphicx}
\usepackage{hyperref}
\hypersetup{
 unicode=false,			
 pdftoolbar=true,		
 pdffitwindow=false,		
 pdfstartview={FitH},		
 pdfcreator={pdflatex},		
 pdfproducer={vim},		
 pdfkeywords={quantum chemistry} {chemical bond} {electron-deficit bonds} {quantum correlations} {entanglement} {multiorbital correlations} {diborane} {C-Be-C complex}, 
 pdfnewwindow=true,		
 colorlinks=true,		
 linktoc=page,			
 linkcolor=blue,		
 citecolor=blue,		
 filecolor=blue,		
 urlcolor=blue			
}

\usepackage[pagewise]{lineno} 

\begin{document}

\title{Towards large-scale restricted active space calculations\\ inspired by the Schmidt decomposition}
\date{\today}
\author{Gergely Barcza}
\affiliation{Wigner Research Centre for Physics, H-1525, Budapest, Hungary}
\affiliation{Department of Physics of Complex Systems, ELTE E\"otv\"os Lor\'and University, H-1117, Budapest,
Hungary}
\affiliation{Department of Chemical and Biological Engineering, The University of Alabama, Tuscaloosa, AL-35487, USA}
\author{Mikl\'os Antal Werner}
\affiliation{Department of Theoretical Physics, Institute of Physics,
Budapest University of Technology and Economics,  H-1111
Budapest, Hungary}
\affiliation{MTA-BME Quantum Dynamics and Correlations Research Group,
 H-1111 Budapest, Hungary}
\author{Gergely Zar\'and}
\affiliation{Department of Theoretical Physics, Institute of Physics,
Budapest University of Technology and Economics,  H-1111
Budapest, Hungary}
\affiliation{MTA-BME Quantum Dynamics and Correlations Research Group,
 H-1111 Budapest, Hungary}
\author{Anton Pershin}
\affiliation{Wigner Research Centre for Physics, H-1525, Budapest, Hungary}
\author{Zsolt Benedek}
\affiliation{Wigner Research Centre for Physics, H-1525, Budapest, Hungary}
\affiliation{Department of Chemical and Biological Engineering, The University of Alabama, Tuscaloosa, AL-35487, USA}
\author{{\"O}rs Legeza}
\email{legeza.ors@wigner.hu}
\affiliation{Wigner Research Centre for Physics, H-1525, Budapest, Hungary}
\affiliation{
Fachbereich Physik, Philipps-Universit\"at Marburg, 35032 Marburg, Germany}
\affiliation{
Institute for Advanced Study,Technical University of Munich, Lichtenbergstrasse 2a, 85748 Garching, Germany}
\author{Tibor Szilv\'asi}
\affiliation{Department of Chemical and Biological Engineering, The University of Alabama, Tuscaloosa, AL-35487, USA}

\begin{abstract}
{\bf Abstract:} We present an alternative, memory-efficient, Schmidt decomposition-based description of the inherently bipartite
restricted active space (RAS) scheme, which 
can be implemented effortlessly within the density matrix renormalization group (DMRG) method via the
dynamically extended active space procedure. 
Benchmark calculations are compared against state-of-the-art results of C$_2$ and Cr$_2$, 
which are notorious for their multi-reference character. 
Our results for ground and excited states together with spectroscopic constants demonstrate that
the proposed novel approach, dubbed as DMRG-RAS, which is variational and free of uncontrolled method errors, has the potential to outperfom conventional methods for strongly correlated molecules. 
\end{abstract}

\maketitle

\section{Introduction} 
The accurate theoretical description of molecular electronic structures is provided by the solution of 
the time-independent Schr\"odinger equation.
In the Born--Oppenheimer approximation, the second quantized representation of the 
corresponding  ab initio Hamiltonian~\cite{aszabo82_qchem} reads 
\begin{equation}
\hat{H}=\sum_{ij,\sigma} t^{\phantom\dagger}_{ij}\hat{a}^\dagger_{i\sigma}\hat{a}^{\phantom\dagger}_{j\sigma} + \frac{1}{2} \sum_{ijkl, \sigma\sigma'}V^{\phantom\dagger}_{ijkl} \hat{a}^\dagger_{i\sigma}\hat{a}^\dagger_{j\sigma'}\hat{a}^{\phantom\dagger}_{k\sigma'}\hat{a}^{\phantom\dagger}_{l\sigma}+E_{\rm nuc}
\label{Ham}
\end{equation}
with system specific one- and two-electron integrals, $t^{\phantom\dagger}_{ij}$, and $V_{ijkl}^{\phantom\dagger}$, respectively.
Here,  operators $\hat{a}^\dagger_{i\sigma}$ and $\hat{a}^{\phantom\dagger}_{i\sigma}$ create and annihilate one  
electron with $\sigma$ spin projection in molecular orbital $i \,\in \{1,\dots,L\}$. 
The dimension of the complete associated computational basis, ${\bf \Phi}$, increases exponentially with spin-orbital 
 size $L$, i.e., $4^L$.
\reversemarginpar
Considering available spatial symmetries and  the conservation of particle and spin quantum numbers, the effective rank of the problem still shows a binomial growth for increasing $L$.
Therefore,  the exact solution based on the full configuration interaction (FCI) method on current 
classical computers is restricted up to around 20 orbitals.~\cite{Vogiatzis2017}

\begin{figure}[h!]
 \includegraphics[width=7cm]{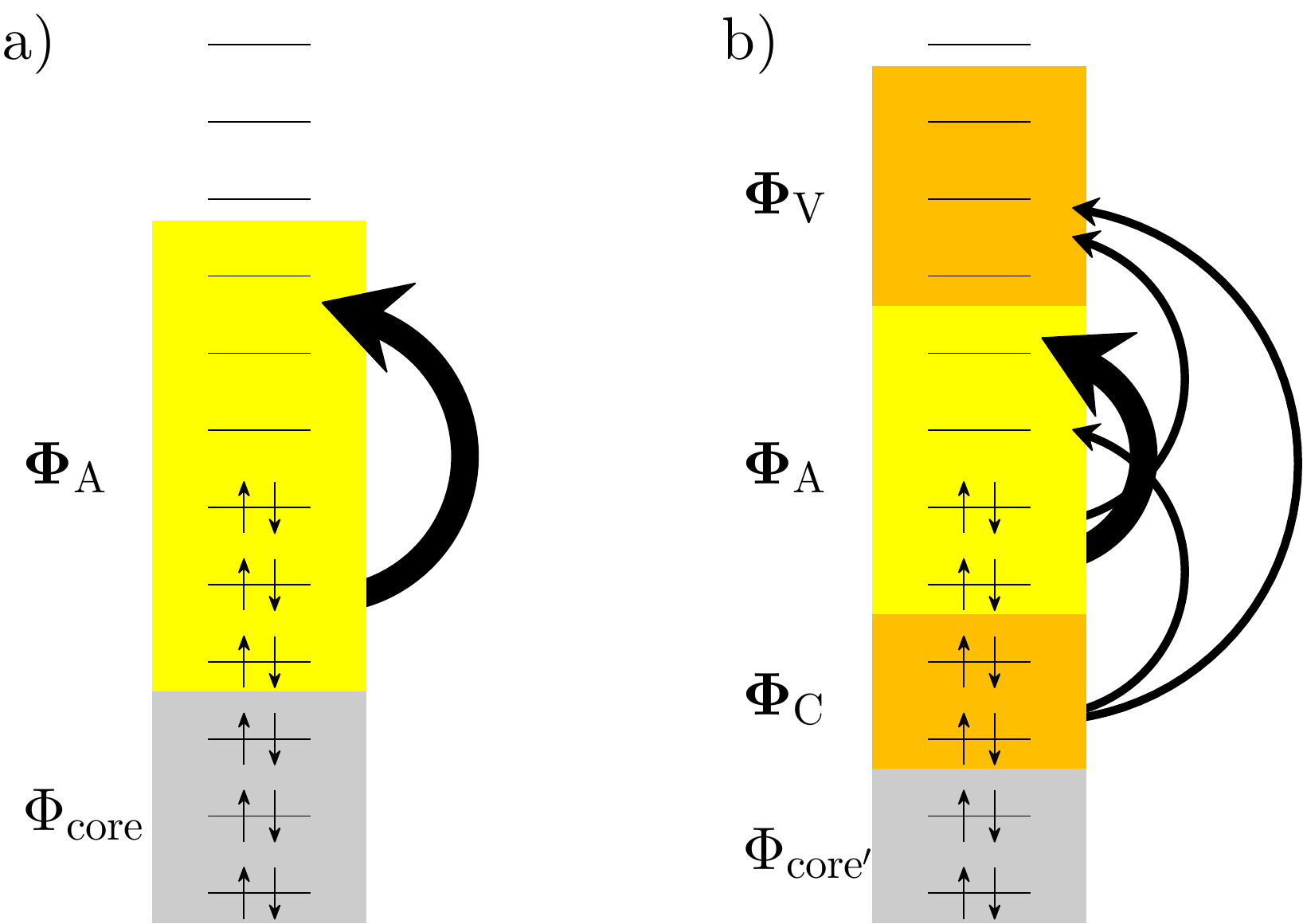}
  \caption{
Schematic illustration of the CAS and RAS concepts. (a) In the CAS scheme 
the occupation of modes can only change for the active orbitals 
(A) while occupation of the core and virtual orbitals is frozen. 
(b) In the RAS scheme, in addition to active orbitals some  virtual (V) 
and  core (C) orbitals can also be excited with restrictions:
the maximal number of particle excitations in these orbitals 
is $r$.
 \label{cas_ras_fig}}
\end{figure}

One protocol that aims to keep the problem of $n_{\rm e}$ electrons computationally feasible is 
the complete active space (CAS) 
scheme~\cite{Roos-1987,Cramer2005,Jensen2006}  which restricts 
the large configuration space to an effective subset spanned by given $L_{\rm CAS}\leq L$  orbitals.
As illustrated in Fig.~\ref{cas_ras_fig}, the protocol classifies the set of canonical molecular orbitals  
into three categories: core, active, and virtual. 
The core and virtual orbitals are kept on the Hartree-Fock level and filled   with 
$n(\Phi_\mathrm{core}) = n_\mathrm{core}$ and zero electrons, respectively.
Note that in this context core orbitals refer to inactive occupied orbitals in general.  
The active orbitals are populated with the rest of electrons, $n_{\rm a}=n_{\rm e}-n_{\rm core}$, minimizing the energy. 
The corresponding determinental space formally reads
\begin{equation}
{\bf \Phi}_{\rm CAS}  \equiv \hat{P}_{\rm CAS} {\bf \Phi } =\{ \Phi_{\rm core}  \Phi_{\rm A}\in {\bf \Phi}  ~|~  n( \Phi_{\rm A})=n_{\rm a}   \}
\label{stdRASstate}
\end{equation}
where the independent conservation of spin channels is not explicitly noted for the sake of brevity. 
Here, $\Phi_{\rm core}$ denotes the string of occupied modes corresponding to the core orbitals while $\Phi_{\rm A}  \in {\bf \Phi}_{\rm A}$  iterates on the set of mode configurations given within the subspace of active orbitals, and the operator $\hat{P}_{\rm CAS}$ projects onto the CAS subspace. 

In practical simulations of large systems, 
a subspace of $L_{\rm CAS}\ll L$ orbitals is selected to focus on the description of states with 
strong (static) correlations, i.e., excited configurations that have largest contribution 
to the ground state.
Even though static correlations characterizes the principal features of the electronic states,  the contribution of intractable number of  excited configurations with small weights,  known as dynamical effects, 
can also be crucial to provide an accurate theoretical description that agrees with experimental data.~\cite{Becke-2013,Benavides-2017}
Although CAS level solutions can be combined with lower-level methods, e.g., self-consistent field theory, 
or coupled cluster method, which aim to provide a balanced description of correlation effects~\cite{Jeppe-2011,Bartlett-2005}
these attempts often have unfavorable drawbacks and limitations.

In the tailored coupled cluster (TCC) method~\cite{Piecuch-1996,Kinoshita_2005,Hino_2006,Veis-2016}, for example, the full orbital space is 
split into two parts, CAS and external (EXT). An accurate solution of the CAS is used to obtain a single reference-like
wave function for a truncated CC approach to recover 
a large fraction of the missing dynamical correlations. 
In spite of the great success of the tailored methods~\cite{Lyakh_2010,Melnichuk_2012,Morchen_2020,Leszczyk_2022}, 
their uncontrollable errors often hinder their application in black-box manner~\cite{Faulstich-2019a,Faulstich-2019}.
In order to tackle these problems within the formalism of coupled cluster theory, besides geminal-based single-reference solutions~\cite{Tecmer_2022}, its computationally challenging multireference extensions have also been  proposed~\cite{Jeziorski_2010,Lyakh_2012,Evangelista_2018,Lischka2018}.
As an alternative approach based on the generalization of the active space scheme~\cite{Fleig_2001,Ma_2011}, the restricted active space (RAS) approach~\cite{Jeppe_1988}, which is practically analogue to the uncontracted multi-reference  configuration interaction~\cite{Lischka2018}, has also gained significant attention~\cite{Casanova_2022}.

In this work, we aim to bring the RAS scheme to a higher level by cross fertilizing it with concept 
of the Schmidt decomposition and with powerful numerical methods based on tensor product approximation, 
like the density matrix renormalization group (DMRG) algorithm~\cite{White-1992b}.
Note that similar concept has also been proposed by grouping orbitals according to various schemes~\cite{Parker_2013,Parker_2014,Nishio_2019}.

Our paper is organized as follows:
In Sec.~\ref{sec:theory} we define the main aspects of the RAS approach and present a 
novel combination of it with Schmidt decomposition.
In Sec.~\ref{sec:ras_method} we discuss implementation details within the framework of the
DMRG method, focusing on the modifications of the 
dynamically extended active space (DEAS) procedure and 
internal basis contractions based on von Neumann entropy.
In Sec.~\ref{sec:results} we demonstrate the efficiency of our new approach for the
C$_2$ and Cr$_2$ molecules.
In Sec.~\ref{sec:con} we summarize the results and present the conclusions of our work.

\section{Theoretical background}
\label{sec:theory}
\subsection{Restricted active space (RAS) approach}
\label{sec:ras}
Conceptually, RAS~\cite{Jeppe_1988,Malmqvist_1990} enhances CAS description by distinguishing two additional subsets of orbitals which define 
the excitable core  (${\bf \Phi}_{\rm C}$) and virtual  (${\bf \Phi}_{\rm V}$) configurational spaces in addition to ${\bf \Phi}_{\rm A}$. 
(Note that the literature~\cite{Malmqvist_1990,Jensen2006} also refers to configurational sets ${\bf \Phi}_{\rm C}$, ${\bf \Phi}_{\rm A}$ and ${\bf \Phi}_{\rm V}$ generated on orbital subsets C, A and V  as RAS1, RAS2 and RAS3 spaces, respectively.)
In the C and V subsets of orbitals we allow a total of $r$ (particle) excitations at most as sketched in Fig.~\ref{cas_ras_fig}. 
 While this restriction in the C and V spaces drastically reduces the computational dimension, 
 the method can still describe the (weak) dynamical correlations and account for their energy contribution.
In the reference  (Slater) configuration the orbitals in space ${\bf \Phi}_{\rm C}$  are completely filled with $n_{\rm c}$ 
electrons, and the active orbitals are occupied by $n_{\rm e}-n_{\rm core'}-n_{\rm c}$ particles,
where $n_{\rm core'}$ is the number of electrons in the low-energy frozen orbitals, ${\bf \Phi}_{\rm core'}$,
that are still kept at the Hartree-Fock level.

The formally considered   $\Phi_{\rm C} \in {\bf \Phi}_{\rm C}$, $\Phi_{\rm V} \in {\bf \Phi}_{\rm V}$ strings of 
modes are defined within the domain of the core and the virtual subspaces and constrained by maximal  
$r$ electronic excitations. 
The associated determinantal space of the RAS scheme we define as

\begin{widetext}
\begin{equation}
{\bf \Phi}_{{\rm RAS}\left[ r \right]}  \equiv \hat{P}_{{\rm RAS}\left[ r \right]} {\bf \Phi}  =\{ \Phi_{\rm core'}\Phi_{\rm C}\Phi_{\rm A}\Phi_{\rm V}  \in {\bf \Phi}  ~|~  n(\Phi_{\rm C}\Phi_{\rm A}\Phi_{\rm V})=n_{\rm e}-n_{\rm core'}~\wedge~ \mathcal{C}\left(n_{\rm c}-n(\Phi_{\rm C}),n(\Phi_{\rm V}), r\right)\},
\label{RAS_space}
\end{equation}
\end{widetext}
where $\mathcal{C}\left(n_{\rm c}-n(\Phi_{\rm C}),n(\Phi_{\rm V}), r\right)$ denotes a condition for  the allowed number of holes in the C orbitals, $n_{\rm c}-n(\Phi_{\rm C})$, and for the  permitted number of electrons in the V virtual space, $n(\Phi_{\rm V})$, according to restrictions based on $r$.
In particular, it is typical to constrain the $\Phi_{\rm C}$ and $\Phi_{\rm V}$ occupations  by requiring
$n_{\rm c}-n(\Phi_{\rm C})\leq r$ and $n(\Phi_{\rm V})\leq r$ conditions simultaneously.
One can introduce even stronger constrains, e.g., $n_{\rm c}-n(\Phi_{\rm C}) + n(\Phi_{\rm V})\leq r$ which  reduces further the electron transfer between C and V spaces.
Note that in the following we use the latter condition and set $r=2$.
Correspondingly, for the sake of compactness, we refer to the presented RAS-based method as DMRG-RAS.

The eigenvalue problem of the corresponding Hamiltonian projected from 
Eq.~\eqref{Ham} by $\hat{P}_{{\rm RAS}\left[ r \right]}$ is written as
\begin{equation}
\hat{H}_ {{\rm RAS}\left[ r \right]}  \Psi_{\alpha}= \hat{P}_{{\rm RAS}\left[ r \right]}  \hat{H} \hat{P}_{{\rm RAS}\left[ r \right]}  \Psi_{\alpha}= E_{\alpha}  \Psi_{\alpha}\;,
\label{RAS_Ham}
\end{equation}
where $\Psi_{\alpha}\in {\bf \Phi}_{{\rm RAS}\left[ r \right]} $ is the $\alpha$-th many-body eigenstate of the projected Hamiltonian.

Considering a problem characterized by $L_{\rm A},~L_{\rm C},~L_{\rm V}$ number of active, 
restricted core and virtual spin-orbitals, respectively, at  RAS[$r$] level of theory, 
the dimension of the associated Hilbert space scales as 
$\mathcal{O}( 4^{L_{\rm A}} (L_{\rm C}+L_{\rm V})^r)$.
Hence, diagonalizing large systems is challenged by 
the memory footprint of the Hamiltonian.
In order to overcome this limitation, one viable strategy, proposed by  the direct-CI 
approach,~\cite{Roos1977,Olsen1990}  is to keep only slices of the matrix 
in memory during iterative diagonalization~\cite{Davidson1975}.

\subsection{Schmidt decomposed RAS} 

As an alternative approach to the problem, hereby we propose a novel implementation of the RAS 
concept which takes advantage of the underlying direct product structure of the 
retained RAS configuration space.
Namely, according to Eq.~\eqref{RAS_space}, the computational basis schematically 
consists of products of configuration states of set 
${\bf \Phi}_{\rm A}$ and set ${\bf \Phi}_{\rm CV}^{(r)}\equiv \{ \Phi_{\rm C}\Phi_{\rm V} \in{\bf \Phi}_{\rm C}  {\bf \Phi}_{\rm V} | n_{\rm c}-n(\Phi_{\rm C})+ n(\Phi_{\rm V} )\leq r\}$. 
It is worth to rewrite also the eigenvalue problem of Eq.~\eqref{RAS_Ham} to emphasize
the product structure, 
\begin{eqnarray}
\hat{H}_{\rm RAS\left[ r \right]}& =&\sum_{i} \hat{O}^{(i)}_{\rm A} \otimes \hat{O}^{(i)}_{\rm CV}\nonumber\\
 \Psi_{\alpha} &=& \sum_{\Phi_{\rm A} \in {\bf \Phi}_{\rm A} \Phi_{\rm CV} \in {\bf \Phi}_{\rm CV}^{(r)} } C_{ \Phi_{\rm A}\Phi_{\rm CV}}^{(\alpha)} \Phi_{\rm A}\Phi_{\rm CV} \,
\label{RAS_state_Sch}
\end{eqnarray}
using coefficient matrix $C^{(\alpha)}$ and adequately constructed  
operators $\hat{O}^{(i)}_{\rm A}$, $\hat{O}^{(i)}_{\rm CV}$ acting on 
subspaces ${\bf \Phi}_{\rm A}$ and ${\bf \Phi}_{\rm CV}^{(r)} $, respectively.
One major consequence of the bipartite representation~\cite{Horodecki-2009} defined 
in Eq.~\ref{RAS_state_Sch} is that the consecutive states of the iterative 
diagonalization via Davidson or L\'anczos algorithm, 
$\Psi'=\hat{H}_{{\rm RAS}\left[ r \right]}\Psi$,  
are obtained from the accumulated sum of product of matrices as
\begin{equation}
C^{(\alpha)'}=\sum_{i} O^{(i)}_{\rm A} C^{(\alpha)} O^{(i)^T}_{\rm CV}.
\label{Ham_bipart_matrixeq}
\end{equation}
Here, the dimension of the matrices reads as 
$|O_{\rm A}|=\left[\mathcal{O}( 4^{L_{\rm A}}), \mathcal{O}( 4^{L_{\rm A}}) \right]$,  $|C_{\alpha}|=\left[\mathcal{O}( 4^{L_{\rm A}}), \mathcal{O}((L_{\rm C}+L_{\rm V})^r)\right]$, $|O_{\rm CV}|=\left[\mathcal{O}((L_{\rm C}+L_{\rm V})^r), \mathcal{O}((L_{\rm C}+L_{\rm V})^r)\right]$.
Consequently for such a bipartite split, the memory requirement of each operator combination might be comparable 
to the footprint of the sliced-direct-CI method in the regime 
where $\mathcal{O}( 4^{L_{\rm A}})\sim \mathcal{O}((L_{\rm C}+L_{\rm V})^r)$.
Note that the number of operator combinations is a quadratic 
function of $L_{\rm A}$ and $(L_{\rm C}+L_{\rm V})$ for quantum chemical Hamiltonians.
For more technical details of the representation see Refs.~\onlinecite{Xiang-1996,White-1999}.

The active orbitals are represented by $\mathcal{O}( 4^{L_{\rm A}})$ number of states in the exact description. 
In order to reduce the computational demands, a truncated many-body basis set based on the Schmidt representation is proposed in the following.

\section{Numerical procedure}
\label{sec:ras_method}

The Schmidt decomposed RAS representation can easily be adapted in the quantum chemical 
density matrix renormalization group (DMRG) approach~\cite{White-1999,Yanai-2009,Chan-2011,Wouters-2014a,Szalay-2015a,Baiardi2020} 
through the dynamically extended active space (DEAS) procedure~\cite{Legeza-2003b,Barcza-2011} 
without additional programming efforts.
In our implementation, the RAS concept was utilized within the 
Budapest-DMRG program package~\cite{budapest_qcdmrg} using the so called two-site variant of DMRG, 
where the superblock is composed of the updated system block, two orbitals treated exactly and the environment block. 
In our representation, the system block extended with the two intermediate orbitals spans the configuration 
space of the active orbitals, ${\bf \Phi}_{\rm A}$, and the environment block describes the 
restricted excitations on core and virtual orbitals, i.e., the corresponding subspace of 
${\bf \Phi}_{\rm CV}^{(r)}$.
The sweeping direction in the DMRG method can be reversed for any superblock configuration, thus we
perform the sweeping only for the active orbital space and we leave the 
subspace of ${\bf \Phi}_{\rm EXT}\equiv{\bf \Phi}_{\rm CV}^{(r)}$ uncontracted. 
Since the size of the active block is always much smaller than the size of the environment block, partial summations 
to form the DMRG auxiliary operators are carried out for the environment block~\cite{Szalay-2015a}.
The number of basis states used to represent ${\bf \Phi}_{\rm CV}^{(r)}$ is much larger than the dimension of a single orbital, thus
the EXT orbital space kept fixed during the DMRG can also be considered as a super-site with large dimension attached to the right end of the DMRG chain. 

From technical point of view, when the DEAS procedure is used, the left block contains one orbital from orbital set A in the first iteration step while the
Hilbert spaces of the 
two intermediate sites together with the 
environment (right) are constructed from the remaining orbitals from set A and from subspaces of the C and V orbitals. The two intermediate sites are treated without any truncation while the right block basis states are formed with excitation rank up to $r=2$. 
The nonzero matrix elements of the operators acting in the RAS space can be generated following the Slater-Condon rules~\cite{Slater_1929,Condon_1930}.
In the course of the forward sweep the size of the left block increased until the pre-set value of the CAS space is 
reached, and in each increment the left block basis is optimized via Schmidt decomposition~\cite{White-1992b,Szalay-2015a}.
Therefore, the left-block is optimized via its
interaction with the RAS space represented by the right block. At the end of the warmup procedure when the
left block together with the two intermediate sites represent the desired number of A
orbital set and the right block is formed only
from the C and V orbital subspaces,
the DMRG sweeping procedure is reversed.
During the backward sweep the right block is optimized~\cite{White-1992b}. In this embedding method, such sweeping procedure is repeated until convergence is reached,
by keeping the number of retained basis states, also known as bond dimension, under control.
In this work we used both fixed bond dimensions and dynamically adapted values via the 
dynamical block state selection approach (DBSS)~\cite{Legeza-2003b,Legeza-2004b} by setting an a priori error margin
for the quantum information loss $\chi$.
In practice, a lower bound on the bond dimension, $M_{\rm min}$, is also enforced to boost convergence.

Note that RAS scheme can also be combined either with orbital optimization in a self-consistent manner or with perturbation theory to further improve the accuracy of the model.~\cite{Malmqvist_1990,Malmqvist-2008}
In our proof of concept implementation of the DMRG-RAS scheme we demonstrate in Sec.~\ref{sec:results} that high-quality solutions can be obtained even without such corrections just by treating large enough completely correlated active spaces. 
In addition, in a recent work~\cite{Friesecke-2022} some of us have proved and demonstrated by large-scale DMRG calculations that the DMRG-RAS energies obtained for increasing A spaces can be extrapolated to the FCI solution within chemical accuracy.

\section{Results and Discussion}
\label{sec:results}
In the following,  as proof of principle, various test calculations are presented for the strongly correlated C$_2$ and Cr$_2$ molecules. 
Ground state energies obtained with different standard quantum chemical methods are compared to the energy calculated by the RAS theory restricted 
up to two electron excitations ($r=2$) denoted as DMRG-RAS in the following.  

The Hartree--Fock electronic structure, multi-reference configuration interaction (MRCI), complete active space self-consistent field (CASSCF) and CASSCF referenced $n$-valence electron perturbation theory (NEVPT2) results are generated using  quantum chemical program suite ORCA~\cite{Neese-2012}. 
High-level configuration interaction (CI) and coupled cluster expansion (CC) solutions as well as Hamiltonian matrix elements, required for the Budapest-DMRG code~\cite{budapest_qcdmrg}, are obtained by MRCC~\cite{mrcc2020,mrcc2020b}.

\subsection{Proof of principle calculations - carbon dimer}
\label{sec:C2_ccpvtz}
 \reversemarginpar
The C$_2$ molecule is modeled at $d=1.25$~\AA{} bond length  in frozen core cc-pVTZ basis~\cite{Dunning-1989} by  correlating 4 electron pairs on 58 spatial-orbitals.
In order to study the convergence of the RAS scheme
towards the FCI energy  as a function of gradually increasing subset A,  a warmup DMRG calculation with DEAS is performed without truncation ($\chi=0$) where the number of the completely correlated A orbitals is  increased in each DMRG microiteration step.
Note that in this benchmark, one molecular orbital (MO) with increasing energy is added to subspace A
in each microiteration
while the rest of the valence space is treated on the restricted level of theory permitting only at most two electron excitations from the reference HF configuration.

\begin{figure}[!t]
   \includegraphics[width=8cm]{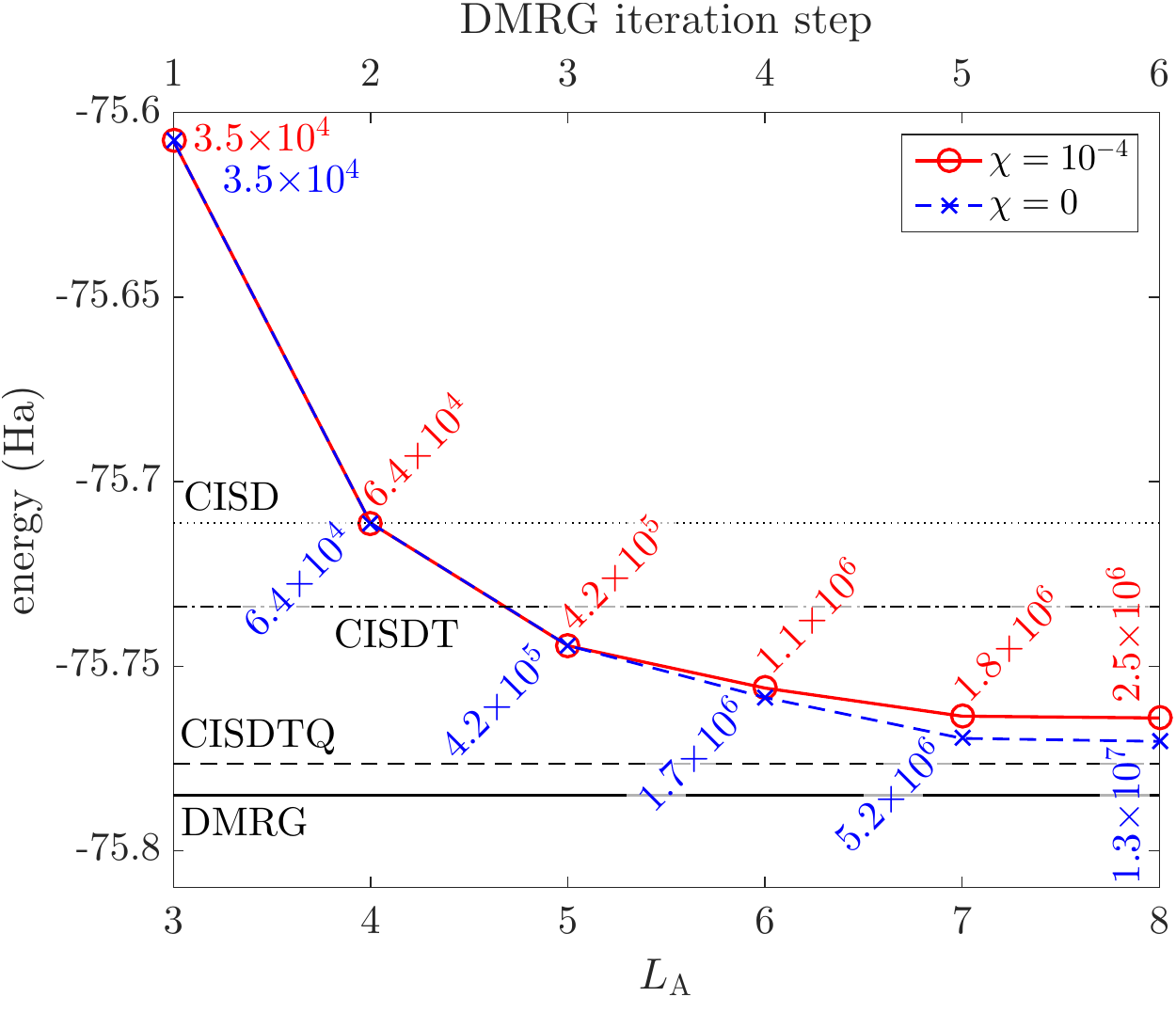}
\vskip -0.3cm
  \caption{Ground state energy of C$_2$ in frozen-core cc-pVTZ basis obtained by DMRG-RAS   after the first half sweep  for increasing 
  $L_{\rm A}$ subspace using the  DBSS truncation scheme. 
  Dimension of the corresponding RAS configurational space is also shown besides the data points in the figure.
  The colored lines are guide to the eye.
  For comparison, reference CISD, CISDT, CISDTQ and large-scale DMRG energy levels are visualized by straight dotted, 
  dashed-dotted, dashed and solid black lines, respectively. 
  For more details of the computational test see Tab.~\ref{tab:C2_ccpvtz}
 \label{fig:C2_vs_LA}}
\end{figure}

In  Fig.~\ref{fig:C2_vs_LA} exact DMRG-RAS energies obtained for active spaces of increasing size ranging from   $L_{\rm A}=3$ to 8 are presented in blue, where the character of the 8 MOs in increasing energy order is identified as $\sigma_{2s}$, $\sigma_{2s}^*$, $\pi_{2p_y}$, $\pi_{2p_z}$,   $\sigma_{2p_x}$, $\pi_{2p_y}^*$, $\pi_{2p_z}^*$, $\sigma_{2p_x}^*$. (The latter notation of $x$, $y$, $z$ directions assumes that both C atoms are located in the $x$ axis).
  In the first DMRG microiteration performed on the active space of $L_{\rm A}=3$ containing only occupied orbitals, i.e., $\sigma_{2s}$, $\sigma_{2s}^*$ and $\pi_{2p_y}$,  the DMRG-RAS solution corresponds to the CISD approximation by allowing up to two particle transfer between C and V by and applying the loose $\mathcal{C}$ constrain defined in Sec.~\ref{sec:ras}.
 Note that these equivalent solutions follow from the additional two-electron restriction on the space A formed purely of occupied orbitals.
Applying the more restrictive constrain on the C and V spaces, a higher energy is obtained as shown in Fig.~\ref{fig:C2_vs_LA}.
For $L_{\rm A}=4$, where the restricted space includes exclusively virtual MOs, the CISD energy is predicted for both restriction constrains.
We find that the energy drops significantly for $L_{\rm A}=5$ where the lowest lying unoccupied molecular orbital of $\sigma_{2p_x}$ character is added to ${\bf \Phi}_{\rm A}$ configurational space which also indicates its strongly correlated character.
The higher lying virtual orbitals incorporated in  $L_{\rm A}=6,7,8$  active spaces, which all have positive orbital energy, affect less drastically the DMRG-RAS level of solution according to the many-body energy and the wave-function analysis.

\begin{table}[!t]
\setlength\extrarowheight{2pt}
  \centering
\begin{tabular}{l|c|c}
\hline
 \hline
method & energy (Ha)& $\Delta_{\rm E}$ (\%) \\
 \hline
HF & -75.4013     & ~~0.0\\
CASCI(8)          & -75.5485 & ~38.4\\
CASSCF(8)          &  -75.6386 & ~61.8 \\
  \hline
CISD             & -75.7112 & ~80.8\\
CISDT            & -75.7339 & ~86.7\\
CISDTQ           & -75.7765 & ~97.8\\
MRCI(8)+Q        & -75.7808 & ~99.0\\ 
CCSD$^a$         & -75.7496 & ~90.8\\
CCSD(T)$^a$      & -75.7832 & ~99.5\\
CCSDT$^a$        & -75.7810 & ~99.0\\
CCSDTQ$^a$       & -75.7845 & ~99.9\\
NEVPT2(8)$^a$     & -75.7540 & ~91.9\\
DMRG-RAS(8,$\chi=10^{-4}$)$^b$ &  -75.7640 & ~94.5 \\
DMRG-RAS(8,$\chi=0$)$^b$       & -75.7704 & ~96.2 \\
 \hline
DMRG-RAS(8,$M=5051$)           & -75.7704 & ~96.2\\
DMRG-RAS(8,$M=500$)            & -75.7694 & ~95.9\\
DMRG-RAS(8,$\chi=10^{-4}$)     & -75.7703 & ~96.2\\
DMRG-RAS(14,$M=500$)           & -75.7753 & ~97.5\\
DMRG-RAS(14,$\chi=10^{-4}$)    & -75.7792 & ~98.5\\
DMRG-RAS(14,$\chi=10^{-6}$)    & -75.7809 & ~99.0\\
DMRG-RAS(18,$M=500$)           & -75.7768 & ~97.8\\
DMRG-RAS(18,$\chi=10^{-4}$)    & -75.7795 & ~98.6\\
DMRG-RAS(18,$\chi=10^{-6}$)    & -75.7836 & ~99.6\\
 \hline
DMRG(58,$M=500$)                 & -75.7779 & ~98.1\\
DMRG(58,$\chi=10^{-6}$)          & -75.7849 & ~99.9\\
DMRG(58,$M=4096$)                & -75.7850 & 100.0\\
 \hline
 \hline
 \end{tabular}
\caption{Ground state energy of C$_2$ in cc-pVTZ basis 
using frozen-core approximation computed by various post-HF methods which correlate 8 electrons. For the active space methods the number of active orbitals are given  in parenthesis while for the DMRG results  the applied truncation scheme is also denoted. 
A series of standard DMRG  calculations performed on the complete space of 58 orbitals is also included. Standard DMRG performed with retained 4096 block states provides the reference correlation energy. 
$\Delta_{\rm E}$ stands for the percentage of correlation energy retrieved by the method.
 The superscript  $a$ denotes the non-variational solutions.
 The superscript $b$ corresponds to energy obtained by the first half DMRG sweep, i.e., at the end of the DMRG warmup procedure. 
 } 
  \label{tab:C2_ccpvtz}%
\end{table}%
By applying the DBSS truncation scheme,  the exact ($\chi=0$) RAS-space results 
can be  approximated in order to reduce computational costs as depicted in Fig.~\ref{fig:C2_vs_LA}  for truncation threshold $\chi=10^{-4}$ and keeping $M_{\rm min}=64$.
In particular, for $L_{\rm A}=8$  the $M_{\rm exact}=4096$ states of the system block are represented by $M=691$ many-body states and accordingly the total Hilbert space of the problem is also reduced by a factor of five.
The introduced truncation yields an energy discrepancy in the range of 0.0 to 0.013 for increasing L$_{\rm A}$ (see also  Tab.~\ref{tab:C2_ccpvtz}). 

Nevertheless, this error can be eliminated by performing DMRG sweeps restricted to the completely correlated orbital set $A$.
As a proof-of-principle, retaining  $M=5051$ block states through the DMRG backward sweep the exact DMRG-RAS energy  can  be accurately reproduced by reaching the left turning point of the sweeping.
Retaining fixed number of block states, $M=500$, while  the rank of the Hamiltonian in the effective subspace  is two orders of magnitude smaller than the uncontracted RAS problem, the exact result can be approached within an error of 1~mHa. 
Utilizing the DBSS procedure with $\chi=10^{-4}$ and $M_{\rm min}=64$ during the DMRG sweeping, in line with expectations,  after a series of DMRG sweeps we get a converged energy with a marginal  error of 0.1~mHa.

As a comparison, reference results of standard post-HF methods are also compiled in Tab.~\ref{tab:C2_ccpvtz} and illustrated in Fig.~\ref{fig:C2_vs_LA}.
We find that although CASCI(8,8) result can be significantly lowered by SCF orbital optimization, it is still substantially higher than cheap CISD approximation due to its poor description of dynamical correlation effects. 
Modeling the problem at CISD, CISDT and CISDTQ level of theory, the retrieved correlation energy with respect to the 
reference DMRG correlation energy (obtained using $M=4096$ and 58 orbitals) increases as 80.8\%, 86.7\%, 97.8\%, respectively.
Comparing to the single-reference CISD and CISDT solutions, DMRG-RAS yields higher resolution of correlations by construction already for $L_{\rm A}=5$ setup.
On the other hand, setting against the standard DMRG energy, which considers all possible excitations without any restrictions, the DMRG-RAS solution is off by 0.022 Ha for $L_{\rm A}=8$ recovering 96.2\% of the total correlation energy. 
Notable that CISDTQ provides practically exact solution retrieving 97.9\% of the correlations. 
In Tab.~\ref{tab:C2_ccpvtz}  CC results of various complexity are  presented as well.
The data clearly demonstrates that the DMRG-RAS approach computed at modest active space size might not only outperform standard single-reference CI expansions but also cheap CC solutions, note that DMRG-RAS provides lower energy than CCSD already for $L_{\rm A}=6$. 
Furthermore,  NEVPT2 results are also  outperformed using DMRG-RAS level of theory. 
Even though the variational RAS based description is computationally more demanding compared to perturbation theory, 
it might provide more-reliable estimate for strongly correlated problems.

Increasing the size of completely correlated space A, the RAS energy decreases at a price of increasing computational efforts.
As a demonstration, in Tab.~\ref{tab:C2_ccpvtz} we also present DMRG-RAS energies obtained for A subspaces extended by virtuals selected according orbital energy.
Describing sizeable active spaces, the basis truncation is already essential to keep the computations feasible.
Retaining $M=500$, we find that extending the 8 orbital active space with MOs 
of character $\sigma_{3p_x}$, $\pi_{3p_y}$, $\pi_{3p_z}$, $\sigma_{3s}$, $\pi_{3p_y}^*$ and $\pi_{3p_z}^*$, yielding $L_{\rm A}=14$ in total, the CISDTQ energy is recovered with an accuracy of 1~mHa.
Applying the more precise DBSS truncation method with $\chi=10^{-6}$,  already CCSDT solution can be reached within 0.1~mHa.
Furthermore, extending the active space to $L_{\rm A}=18$ orbitals with 
$\sigma_{3s}^*$, $\delta_{3d_{yz}}$, $\delta_{3d_{y^2-z^2}}$ and $\sigma_{3p_x}^*$ MOs, already CCSDTQ prediction can be retrieved within chemical accuracy.

\begin{table}[!b]
\setlength\extrarowheight{2pt}
  \centering
\begin{tabular}{l|c|c|c|c|c|c}
\hline
 \hline
 \multirow{2}{*}{method} & \multicolumn{6}{ c }{$d$ (\AA)}\\
 \cline{2-7}
 &    1.25&  1.75&  2.25&  2.75&  3.25&  5.00\\
\hline
CISD   &   -0.711& -0.568& -0.517& -0.440& -0.394& -0.368\\
CISDT   &   -0.734& -0.593& -0.541& -0.473& -0.451& -0.452\\
CISDTQ &   -0.777& -0.660& -0.591& -0.553& -0.547& -0.550\\
MRCI(8)+Q   &   -0.781& -0.674& -0.599& -0.570& -0.564& -0.562\\
CCSD     &    -0.750& -0.613& -0.559& -0.513& -0.504& -0.334\\
CCSD(T) &   -0.783& -0.654& -0.595& -0.581& -0.615& -0.498\\
CCSDT   &   -0.781& -0.654& -0.594& -0.576& -0.582& -0.424\\
CCSDTQ  &   -0.785& -0.651& -0.601& -0.573& -0.569& -0.570\\
CASSCF(8)   &  -0.639& -0.514& -0.444& -0.417& -0.413& -0.412\\
NEVPT2(8)   &   -0.754& -0.662& -0.576& -0.539& -0.534& -0.532\\
\hline
RAS(8)$^a$       &   -0.769& -0.682& -0.587& -0.558& -0.551& -0.551\\
RAS(18)$^a$       &   -0.777& -0.688& -0.593& -0.560& -0.555& -0.556\\
RAS(18)$^b$  &  -0.784& -0.695& -0.601& -0.570& -0.563& -0.561\\
 \hline
 \hline
 \end{tabular}
\caption{ 
Potential energy surface  of the C$_2$ molecule  calculated for various interatomic distances $d$.
The ground state energies obtained by different post-HF methods, measured in Ha, are shifted by 75.00 Ha and rounded to third digit resulting an error comparable to chemical accuracy.
In case of superscript $a$  DMRG-RAS calculations were performed with fixed $M=500$, in case of superscript $b$ the DBSS scheme with $\chi=10^{-6}$ was used.
}
  \label{tab:C2_ccpvtz_r}%
\end{table}%

\subsubsection{Stretching the geometry}
Next, we show that our approach is not restricted only to the equilibrium geometry but it can be applied through the whole potential energy surface (PES). 
Ground state energy of  C$_2$  obtained by various post-HF methods  for various interatomic distances, $d$, together with DMRG-RAS solutions 
are collected in Tab.~\ref{tab:C2_ccpvtz_r}.
It is clearly visible that even a very cheap DMRG-RAS(8,$M=500$) provides lower energy bounds compared to CCSD along the whole PES, while DMRG-RAS(18,$M=500$) is already comparable to CCSDT level of theory.
Increasing the accuracy of the RAS(18) calculation by  fixing the truncation error to $\chi=10^{-6}$, good agreement is found with CCSDTQ calculated PES except for the most correlated setup at $d=1.75$~\AA{}, where DMRG-RAS predicts significantly lower energy.
Note that MRCI(8) with Davidson Q correction consistently confirms the trends of the DMRG-RAS(18,$\chi=10^{-6}$) solution.
In fact, MRCI(8)+Q predicts energy within chemical accuracy  for large separations, $d\geq 2.75$~\AA,  while it provides $0.002-0.02$~Ha higher energies for moderate interatomic distances, $d<2.75$~\AA.
Note also that DMRG-RAS method performed on a given A space is variational, thus increasing $M$ or lowering $\chi$ provides a monotonic and stable convergence and even the cheap  DMRG-RAS(8,$M=500$) gives a smooth PES. 
Therefore, for a fixed bond length no oscillations are experienced like in CC for $d=3.25$~\AA.

Based on the PES, experimentally measurable quantities  - such as equilibrium bond length, $r_{\rm e}$ harmonic vibration frequency, $\omega_{\rm e}$, and dissociation energy, D$_{\rm e}$, - can also be derived. 
Scanning the PES around the equilibrium,
in the range of  1.235~\AA{} to 1.275~\AA, 
we found that our numerical data excellently fits the harmonic approximation.
The DMRG-RAS dissociation energy, D$_{\rm e}$,  is derived from the energy of two  carbon atoms computed at separation $d=1000~\AA$.
The DMRG-RAS results together with benchmark results obtained by  
various others methods are presented in Tab.~\ref{tab:C2_parameters}.
We find that the DMRG-RAS results depend slightly on the active space size, however, for $L_{\rm A}=18$ the high-level quantum Monte Carlo based FCIQMC reference data are reproduced already up to three digits.
Note that experimental data, which is also included for completeness, is not expected to be precisely predicted by the numerical results owing to the limitations of the finite basis set and to the frozen core approximation.
\begin{table}[!t]
\setlength\extrarowheight{2pt}
  \centering
 \begin{tabularx}{0.48\textwidth}{ l|c|c|c }
\hline
\hline
Method & ~$r_{\rm e}(\AA)$~& ~$\omega_{\rm e}$(cm$^{-1}$)  ~& ~D$_{\rm e}$(kcal/mol)  \\
\hline
CCSD~\cite{Leszczyk_2022}         &  1.240 & 1915 & 156.6 \\
DMRG(8)-tCCSD~\cite{Leszczyk_2022}    &   1.249 & 1885 &  146.3 \\
MRCI~\cite{Peterson_1995}     &    1.252 & 1840 & 140.4 \\
FCIQMC~\cite{Booth_2011} &     1.253 & 1835 & 135.4 \\
\hline
DMRG-RAS(8)   &     1.255 & 1821 & 140.9 \\
DMRG-RAS(18)   &     1.253 & 1839 & 135.8 \\
\hline
Experiment~\cite{Huber_1979} & 1.243  &  1855 & 147.8 \\
\hline
\hline
 \end{tabularx}
\caption{
Spectroscopic constants  for the dissociation of C$_2$ at different levels of theory in cc-pVTZ basis. 
Parameters $r_{\rm e}$, $\omega_{\rm e}$ and D$_{\rm e}$ denote equilibrium bond distance, harmonic vibration frequency and dissociation energy, respectively.}

  \label{tab:C2_parameters}%
\end{table}%

Finally, recalling that the linear  configurational interaction theory  applied for a single reference is notorious for the lack of size-consistency~\cite{Pople_1987}. Here, we also investigate this open problem for the RAS description which essentially provides a multi-referenced CI solution.
As a test, we compared the DMRG-RAS(18) energy obtained for C$_2$ separated by 1000~\AA, $E=-75.5606$~Ha, and the twice of the DMRG-RAS energy for a single C atom keeping 9 orbitals in the active space, $E=2\times  -37.7808= -75.5616$~Ha, which yields a discrepancy below chemical accuracy.
Note that the accuracy of size-consistency in the RAS calculations can be controlled by the active space selection and the choice of basis, furthermore it can also be improved a posteriori by applying the Davidson correction~\cite{Langhoff1974} or by implementing the  quadratic CI theory~\cite{Pople_1987}. A more detailed analysis, also for more complex systems, is under progress and will be part of our subsequent work. 

\subsubsection{Convergence in large basis set}
\label{sec:C2_ccpvqz}
\begin{figure}[!t]
  \includegraphics[width=9.5cm]{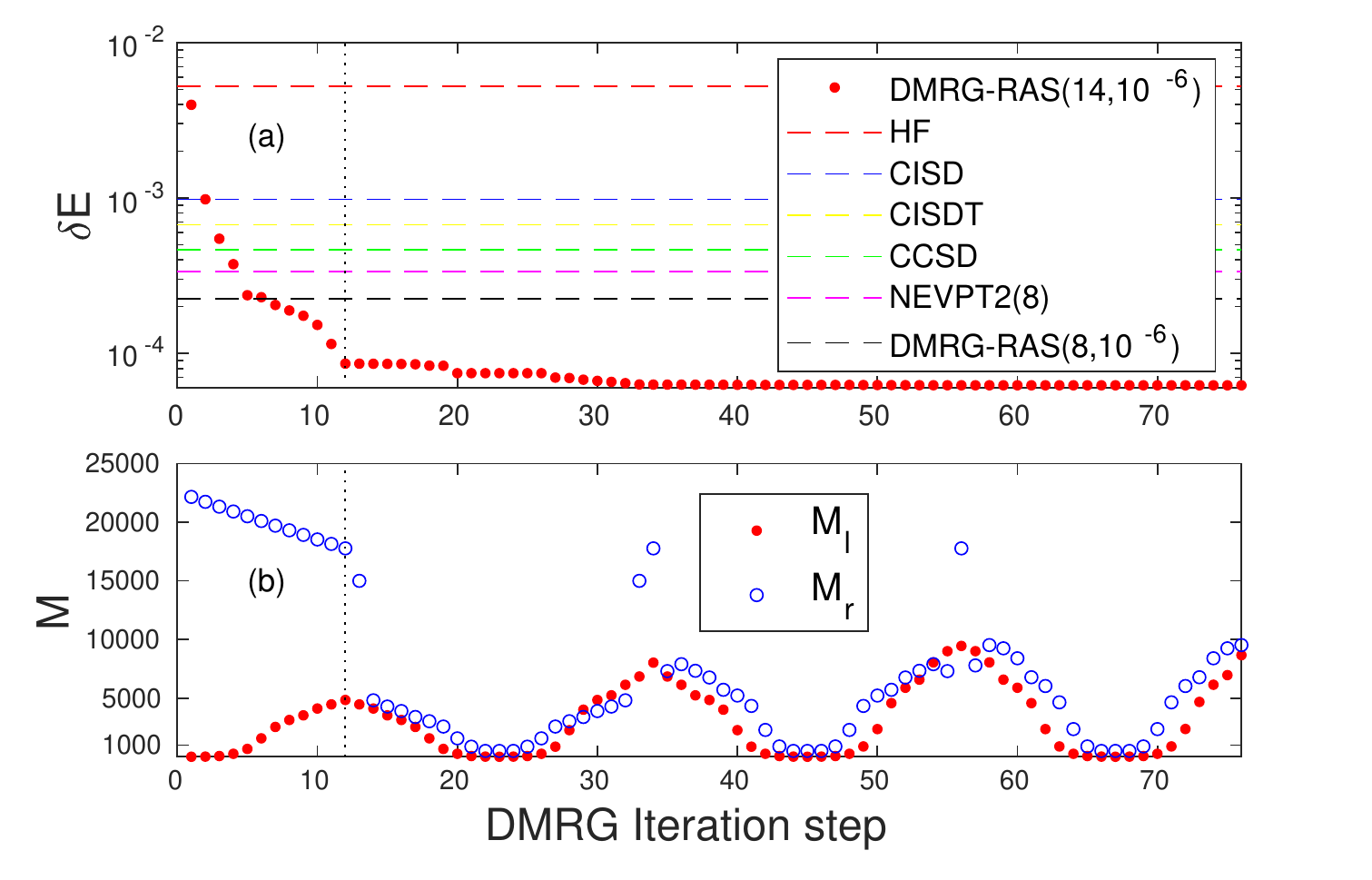}
\vskip -0.3cm
  \caption{
(a) Relative error,
$\delta E$, of the ground state energy with respect to the CCSD(T) reference 
energy, $E=-75.8007$(Ha), obtained by various standard methods and 
by the RAS-SD-DMRG for $L_{\rm A}=8,14$ and $\chi=10^{-6}$ and 
for cc-pVQZ basis corresponding to $L=108$ orbitals.
(b) For $L_{\rm A}=14$ the convergence profile
as a function of DMRG iteration steps in terms of the corresponding number of the retained DMRG block 
states, $M$.
Vertical dotted line indicates the last step of the first half DMRG sweep,
 i.e., at the end of the DMRG warmup procedure. 
}  
\label{fig:C2_ccpvqz}
\end{figure}

 Numerical results obtained for C$_2$ at separation $d=1.25$~\AA{} modeled in cc-pVQZ basis set with frozen 
cores are shown in Fig.~\ref{fig:C2_ccpvqz} and also summarized in Tab.~\ref{tab:C2_ccpvqz}.
Considering the extensive number of orbitals, $L=108$, the analysis 
is restricted to the computationally less demanding 
methods which confirms trends found in the smaller basis calculations.
\begin{table}[!b]
  \centering
\setlength\extrarowheight{2pt}
\begin{tabular}{l|c}
\hline
\hline
method & energy (Ha) \\
 \hline
HF & -75.40561 \\
  \hline
CISD &  -75.7264\\
CISDT  &  -75.7497\\
CCSD$^a$  & -75.7656\\
CCSD(T)$^a$ &  -75.8007\\
NEVPT2(8)$^a$ &  -75.7753\\
DMRG-RAS(8,$\chi=10^{-6}$)$^b$ & -75.7749\\
  \hline
DMRG-RAS(8,$M=500$)           & -75.7828 \\
DMRG-RAS(8,$\chi=10^{-6}$)    & -75.7837 \\
DMRG-RAS(14,$\chi=10^{-6}$)   & -75.7971  \\
 \hline
 \hline
\end{tabular}
\caption{Similar test calculations as in Tab.~\ref{tab:C2_ccpvtz}, but for cc-pVQZ basis corresponding to $L=108$ orbitals.
The $L_{\rm A}$ number of completely correlated orbitals used in NEVPT2 and DMRG-RAS calculations are noted in parenthesis. For DMRG-RAS calculations the applied truncation criteria is also given.
The superscript $a$ and $b$ denote the non-variational solutions and the energy obtained by the first half DMRG sweep (warmup procedure), respectively.} 
  \label{tab:C2_ccpvqz}%
\end{table}%

Most notably,  DMRG-RAS applied for active spaces of moderate size is already capable to  outperform CISDT, CCSD and NEVPT2 energies after the first half DMRG sweep, see Tab.~\ref{tab:C2_ccpvqz} for $L_{\rm A}=8$ and Fig.~\ref{fig:C2_ccpvqz} for $L_{\rm A}=14$.
By carrying out sweepings for $L_{\rm A}=8$ with fixed $M=500$ block states or by fixing $\chi=10^{-6}$ the energy can be further lowered significantly. 
In addition, increasing $L_{\rm A}$ to 14 the CCSD(T) reference energy can be reached within chemical accuracy. 
As shown in Fig.~\ref{fig:C2_ccpvqz}b, the tremendous drop in the number of retained block states after the first sweep demonstrates the efficiency of the DMRG basis state selection.
Note that in case of computationally demanding problems, ${\bf \Phi}_{\rm CV}$ can also be included in the DMRG sweeping procedure with further constraints and an optimally selected ${\bf \Phi}_{\rm CV}$ space can be constructed according to quantum information entropies~\cite{Legeza-2003b}.
For more complex systems, DMRG ordering optimization together with orbital optimization to provide more appropriate orbital set A can also be critical as will be shown below.

\subsection{Optimization strategies for more complex systems - case study of chromium dimer}
\label{sec:cr2}
In the following, we study the performance of the DMRG-RAS approach for the strongly correlated chromium dimer 
which multi-reference problem is also
often used to benchmark DMRG-based methods~\cite{Kurashige-2011,Sharma-2012,Veis-2016,Ma-2017,Larsson_2022}.
Calculations are performed for the Cr$_2$ molecule in its equilibrium geometry, $d=1.6788$~\AA, in the cc-pVDZ atomic basis which model the correlation of the twelve $3d 4s$ electrons 
in the space of 68 spin-orbitals
while keeping all other lower-energy electrons frozen.  
The obtained post-HF results for the singlet and triplet lowest lying states
are summarized in Tab.~\ref{tab:benchmark_Cr2}

For the singlet state, we observe the relative low variance of the high-level CC reference data which might 
indicate a reliable convergence to the FCI solution.
In contrast, the computation of the lowest triplet state by CI and CC methods is proven to be more challenging as will be shown below.
Since single reference methods might depend heavily on the chosen reference determinant, we performed independent CC calculations by varying the SCF settings in the MRCC program package.
In Tab.~\ref{tab:benchmark_Cr2}, the lowest CI and CC energies, which were obtained by the default SCF setting, is shown.
Being more sensitive to the reference determinant, we find that the triplet energy drops substantially for increasing model complexity contrary to the singlet case.
Consequently, the singlet-triplet gap, expected to be around 1 eV~\cite{Sharma-2012}, 
is significantly overestimated by the presented CI and CC solutions.
\begin{table}[!t]
\setlength\extrarowheight{2pt}
  \centering
 \begin{tabular}{ l|c|c|c }
\hline
\hline
method & ~singlet (Ha)~ & ~triplet (Ha)~ & gap (eV) \\
\hline
CISD	     &	-2086.5501 & -2085.9127  & 17.34\\
CISDT	     &	-2086.5932 & -2086.2260  &  ~9.99\\
CCSD	     &	-2086.7401 & -2086.0758  & 18.08\\
CCSD(T)    &	-2086.8785 & -2086.5059  & 10.14\\
CCSDT      &	-2086.8675 & -2086.7042  &  ~4.44\\
CCSDTQ	   &	-2086.8689 & -2086.7282  &  ~3.83\\
CASSCF(12) &	-2086.6502 & -2086.6213  &  ~0.79 \\
NEVPT2(12) &	-2087.2744 & -2087.2460  & ~0.77 \\
\hline
RAS(12,$M$=500)$^{a}$ & -2086.5626  &	-2086.5005 & ~1.69 \\
RAS(12,$M$=500)$^{b}$ & -2086.7686  & -2086.7364 & ~0.88 \\
RAS(12,$M$=500)$^{c}$ & -2086.8516	& -2086.8236 & ~0.76 \\
RAS(20,$M$=500)$^{c}$ & -2086.8542  & -2086.8263 & ~0.76 \\
RAS(20,$M$=1000)$^{c}$ & -2086.8646 & -2086.8359 & ~0.78 \\
\hline
\hline
 \end{tabular}
\caption{
Ground state energy of Cr$_2$ in its equilibrium geometry 
obtained by different post-HF methods in cc-pVDZ basis and 
frozen-core approximation.
The correlation of the twelve $4s3d$ electrons are 
included explicitly by all the non-perturbative approaches.
The $L_{\rm A}$ number of completely correlated orbitals used in CASSCF, NEVPT2 and DMRG-RAS calculations are noted in parenthesis.
For DMRG-RAS calculations the number of retained block states $M$ is also given.
Superscripts $a$ and $b$ correspond to DMRG-RAS calculations performed on active orbitals selected from the canonical orbital set around the HOMO-LUMO gap or to the orbital entropy  
profile, respectively.
The superscript $c$ denotes DMRG-RAS results obtained in natural orbital basis, where the completely correlated orbitals were selected according to their chemical relevance reflected by large orbital entropies.
} 
  \label{tab:benchmark_Cr2}%
\end{table}%

Turning the focus to the DMRG-RAS theory, we present several computational strategies which  
have the potential to  improve systematically the convergence of the method.
In the most naive computational scheme, we correlate 12 electrons on the 12 canonical orbitals around the HOMO-LUMO gap in active space A. The RAS theory provides only slightly lower energy than the standard CISD solution indicating the inadequacy of the virtual orbitals selected for subspace A, see DMRG-RAS energies with superscript $a$ in Tab.~\ref{tab:benchmark_Cr2}.
\begin{figure}[!b]
  \includegraphics[width=0.45\textwidth]{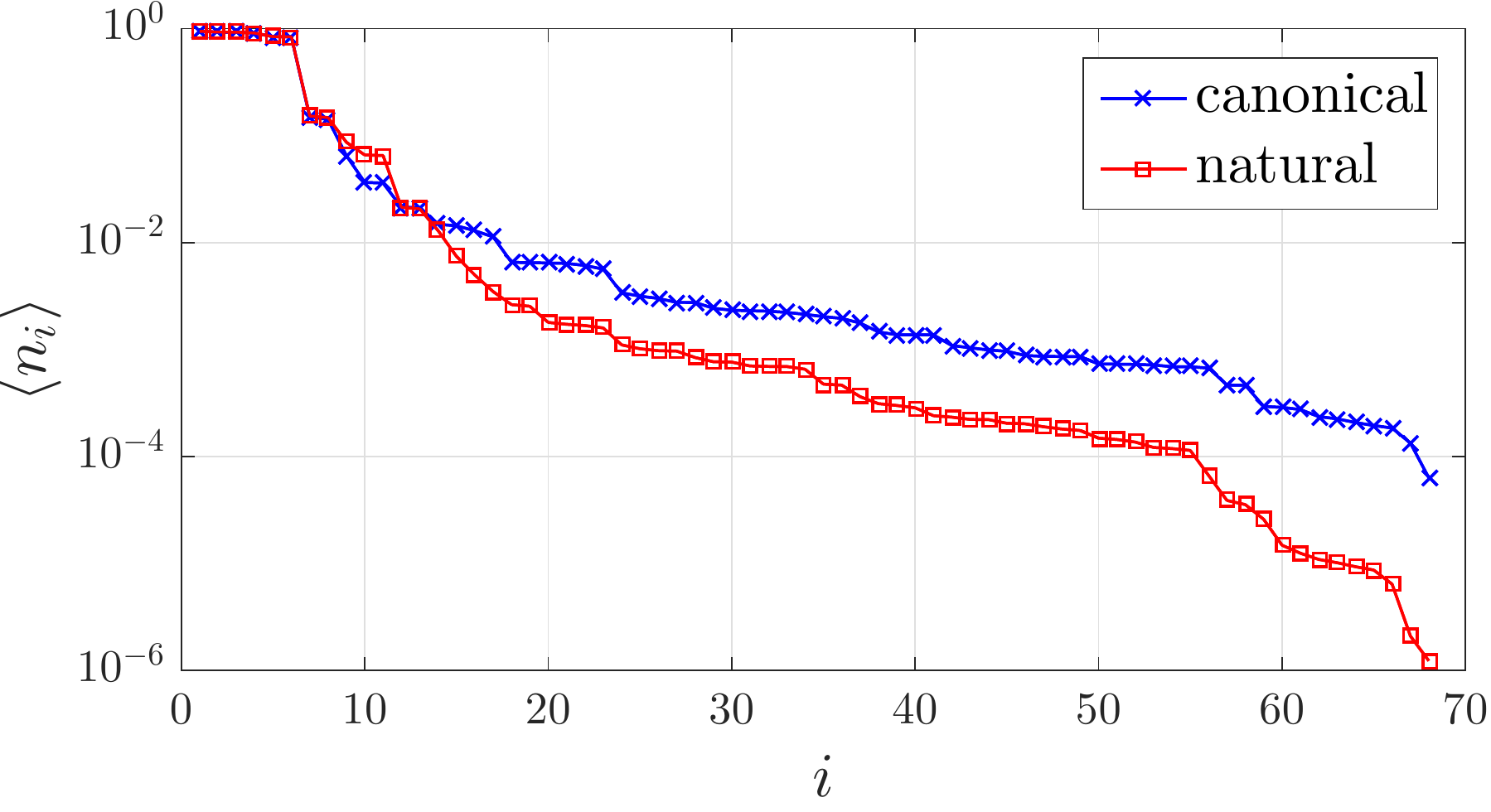}
  \caption{Sorted orbital occupation calculated by a low-accuracy ($M=300$) DMRG for the ground state of the 
chromium dimer obtained in the canonical orbital and in the corresponding natural orbital basis set. 
The solid lines are  guides to the eye.
 \label{fig:NO}}
\end{figure}
Note that completely correlated subset of canonical orbitals is formed of the six bonding orbitals ($\sigma_{3d_{x^2}}$, $\sigma_{4s}$ as well as the twofold degenerate $\pi_{3d}$ and $\delta_{3d}$) and the virtual orbitals with lowest energies.
We find that these MOs are composed of $4s$ and $4p$ atomic orbitals while the chemical intuition suggests that the relevant virtual orbitals are those antibonding MOs which correspond to bonding orbitals featuring $4s$ and $3d$ atomic orbitals.
This assumption is also corroborated by studying the single orbital entropy profile~\cite{Legeza-2003b,Stein_2016} of the molecule to identify the relevant orbitals, which we 
obtained from a low-precision ($M=300$) standard DMRG calculation on the complete space of 68 orbitals.
Accordingly, by choosing the six chemically most relevant antibonding  $4s$ and $3d$  virtual orbitals in subspace A, the DMRG-RAS energy drops drastically and gets even below the CCSD reference energy (see superscript $b$ in Tab.~\ref{tab:benchmark_Cr2}).

In order to further improve the DMRG-RAS energy, we recall that the restricted excitations of the RAS space could only represent those orbitals faithfully whose chemical importance is negligible, i.e., orbitals with occupation number close to zero or two.
Accordingly, by expressing the Hamiltonian in the natural orbital basis based on a low-precision ($M=300$) DMRG calculation, 
orbitals with larger partial occupation are 
concentrated only on a few number of virtual orbitals and the rest of the spectrum has
a much faster decay compared to the one obtained for canonical orbitals, see Fig.~\ref{fig:NO}.
Utilizing the natural orbitals in the DMRG-RAS method with
$L_{\rm A}=12$ active orbitals the CCSDT energy can be recovered up to 0.016 Ha, see RAS data indicated by superscript $c$ in Tab.~\ref{tab:benchmark_Cr2}.
Furthermore, using the same strategy but extending  the completely correlated subspace A by keeping $L_{\rm A}=20$ orbitals, the CCSDT ground state energy is reproduced practically up to chemical accuracy (0.003 Ha).

Repeating the same analysis for the lowest-lying triplet state the DMRG-RAS provides again a stable and monotonic
convergence with increasing $L_{\rm A}$ and $M$ values. Using natural orbitals the calculated gap shows only
very small variance for increasing $L_{\rm A}$ and $M$ values.
It is also remarkable that even though  CASSCF, NEVPT2 and high-level DMRG-RAS report substantially different absolute energies, all these three methods predict energy gap around 0.78 eV which result is also in line with previous findings~\cite{Sharma-2012}.   
Note that the lowest absolute energies of the NEVPT2 solution, which retrieves correlation effects of the core electrons in a perturbative manner, imply the relevance of lower-energy $3s3p$ electrons for a reliable description of the chromium dimer.

\section{Conclusions}
\label{sec:con}
In this work, we proposed a formulation of the restricted active space scheme considering the Schmidt decomposition 
which can be efficiently treated by the density matrix renormalization group method
through the dynamically extended active space (DEAS) procedure without additional programming efforts.
The dimension of the Hilbert space of the uncontracted RAS orbital space can easily be chosen to be in the range 
of a few tens of thousands also taking into account the sparsity of the corresponding matrices. 
The iterative decomposition also provides a possibility to truncate the Hilbert space of the active orbital space 
using quantum information measures. Most importantly, optimization of the CAS space is carried out in the presence
of the EXT space, unlike in case of other embedding methods.
As a proof of concept, we tested the method dubbed DMRG-RAS on the strongly correlated problem of \reversemarginpar
carbon
and chromium dimers 
which is compared to standard state-of-the-art numerical solutions rivaling with respect to accuracy and computational demand.
Our method is variational and provides a stable and monotonic convergence with increasing size of the active space
and DMRG bond dimension not only at equilibrium geometries, but along the whole potential energy surface
even when multireference character gets more pronounced.
When lowest lying excited states are targeted in different quantum number sectors similar properties hold. 
The performance of the DMRG-RAS method can be further boosted by basis optimization and utilization of concepts
of quantum information theory.
The generalization of the presented computational scheme  as a multilayered protocol is straightforward where
most RAS orbitals are treated only up to double excitation level while the more intricate ones are at higher level of theory.
The method is expected to be particularly well-suited to provide a balanced description of static and dynamical  
correlations of ab initio systems with orbital space separable according to its correlation pattern.

While completing our work we got aware of a similar approach~\cite{Larsson-2021} suggesting to attach large sites 
to the two ends of the DMRG chain.
A more detailed analysis of excited states together with improving
the CC approach via DMRG-RAS is part of our current work.

\acknowledgements
This work was supported by the Ministry of Innovation and Technology and the National Research, 
Development and Innovation Office of Hungary (NKFIH) within the  
Quantum Information National Laboratory of Hungary 
and project No.~K134983, No.~TKP2021-NVA-04 and No.~FK-135496.
M.A.W. has also been supported by the \'UNKP-21-4 and NKP-22-5-BME-330 New National Excellence Program of the Ministry for Culture and Innovation from the source of the National Research, Development and Innovation Fund.
The Bolyai Research Scholarship of HAS is greatly acknowledged by G.B.
and M.A.W.
We acknowledge KIF\"U for awarding us access to computational resource based in Hungary.
T.S. would like to thank the University of Alabama and the Office of Information 
Technology for providing high performance computing resources and support that have contributed to these research results. 
This work was made possible in part by a grant of high-performance computing resources 
and technical support from the Alabama Supercomputer Authority.
\"O.L. also acknowledges financial support from the Hans Fischer Senior Fellowship programme
funded by the Technical University of Munich -- Institute for Advanced Study.
The development of DMRG libraries has been supported by the Center for
Scalable and Predictive methods for Excitation and Correlated phenomena (SPEC),
which is funded as part of the Computational Chemical Sciences Program by the U.S.\
Department of Energy (DOE), Office of Science, Office of Basic Energy Sciences,
Division of Chemical Sciences, Geosciences, and Biosciences
at Pacific Northwest National Laboratory.
 
Dedicated to the memory of G\'eza Tichy and J\'anos Pipek.

\end{document}